\documentclass[twocolumn,a4paper,showpacs,showkeys,amsmath,amssymb]{revtex4}
\usepackage{graphicx}

\begin{document}
\title{Observation of free hole gases at ambient conditions}
\author{H.C. Mastwijk}
\email{hennie.mastwijk@wur.nl}
\author{H.J. Wichers}
\author{C. van Dijk}
\author{H.J. Schuten}
\affiliation{Wageningen UR, Bornsesteeg 59, 6708 PD, Wageningen,\\The Netherlands,
p + 31 317 481307, f + 31 317 483011}
\date{29-11-2007, revision 19-04-2009}

\begin{abstract}By studying fluctuations in the electrostatic potential of a single electrode, we were able to perform {\it in situ} electron spectroscopy. Electron exchange processes that occur at the surface of the electrode were triggered by the surface Auger process either using metastable molecular nitrogen or excited state helium at ambient conditions. Ongoing redox reactions were decomposed into their two fundamental half reactions and the reduction and oxidation potentials were detected. We found that redox reactions near the electrode surface are the result of binary interactions of a free electron gas with a free hole gas that occur on a femtosecond time scale. The measured lifetimes of electron-hole recombination processes and the elastic scattering rate of the hole-hole process are in fair agreement with theoretical estimates. The observed asymmetry in the energy distributions of free electron and free hole ensembles suggests that at thermal equilibrium the identity of hole differs from that of an electron vacancy.
\end{abstract}

\maketitle

\section{Introduction}
\label{sec:Intro}
The ultimate goal in quantum chemistry is to achieve control over molecular reactions at a single electron level. The development of diagnostic tools to detect specific electron processes has been a continuous effort for decades \cite{Cagnon,Niehaus,Ohno}. Of practical significance are those tools that can operate under ambient conditions such as {\it in situ} STM and electrochemical devices \cite{Lemay,Heller,Frenken}. 

The physics and chemistry studied in a system that is held under ambient conditions (high particle densities), is fundamentally different from the physics and chemistry of the same system in a vacuum (low densities). Under conditions where a free molecular flow is established, no scattering occurs until a reaction takes place. At high particle density elastic scattering rates are such \cite{Reif}, that a complete thermalisation of the system has occurred by collisions of electrons with molecules in the background, before and after reactions have taken place. As a result, the response of an electron system that is evaluated at high densities proceeds through the response of an ensemble of particles that has reached a thermal equilibrium. This is in contrast to the low density case where the response of the system is evaluated by the response of a single electron at a fixed energy.

Electrochemical devices are based on the detection and quantification of charge transfer. In the analysis of redox reactions the energy required in the process of electron transfer is a measure of the standard redox potential of the specific molecule under investigation. If one is able to quantify electron energies with sufficient resolution it is possible to discriminate between different ongoing redox reactions and resolve the molecular chemistry to the level of single electronic quantum states \cite{Niehaus,Ohno}. In this paper we provide experimental evidence of the possibility of single electrode quantum coulometry as tool to perform {\it in situ} high resolution electron spectroscopy. In essence we present a method which allows the study of isolated half reactions that corresponds to well defined reduction (oxidation) states by a direct count of the number of electron (hole) states and differentiate with respect to their redox potential. An important feature of the method is that measurements are carried out at ambient, nearly field free conditions without the principle requirement of a counter electrode. In order to understand the electronic response of a single electrode subjected to chemical processes at its outer surface, we have re-examined the physics of electron ensembles at the interface of an electrode \cite{Ashcroft} using the principles of surface Auger spectroscopy \cite{Hotop}. We found that neither high vacuum conditions nor electron spectrometers employed in conventional Auger spectroscopy are requisites. The entire experimental set-up was reduced to a single working electrode operated at ambient conditions, directly recording the redox reactions near the interface which were detected as events that cause fluctuations in the electrode potential.

This paper is organized as follows. First, we shall discuss principle features known in quantum physics concerning the measurement of electronic energies and motivate how voltage fluctuations are converted into electron energy spectra. Secondly, we shall provide experimental evidence that electron energy spectra can be derived from fluctuations in the electrostatic potential by studying the surface Auger process invoked by excited state nitrogen and helium when these gases are exposed to a charged metal surface. Finally, we will argue that redox events are the result of electron-hole recombination interactions of a free electron gas and a free hole gas that are omnipresent. 

\section{Theoretical framework} 
\label{sec:Theory}
The electronic wave function of a discrete quantum system \cite{Bransden} is expressed in terms of its eigen states $\psi_i$ by

\begin{equation}
\Psi=c_1\psi_1+c_2\psi_2+...c_n\psi_n
\label{wavefunction}
\end{equation}      					

The expectation value of the energy is given in terms of its eigen energies $E_i$ by \cite{Bransden}  

\begin{equation}
\left\langle E\right\rangle=c_1^2E_1+c_2^2E_2+...c_n^2E_n=P_iE_i
\label{expectation value}
\end{equation}
  
where $P_i$ is the probability of finding the system at a specific eigen energy $E_i$. It is important to realize that the expectation value is not a true observable but a derived quantity. Certainly, a detector can be configured to perform an averaging procedure to read an actual outcome of magnitude $\left\langle E\right\rangle$ but the only irreducible observables (actual outcome of an measurement) are the eigen energies $E_i$. As an example, consider playing dice. The outcome of each throw is an element of the set (1,2,3,4,5,6). After collecting a large number of outcomes the expectation value, or arithmetic mean, approaches the value $3 \frac{1}{2}$. However, the expectation value is never observed as an actual outcome.  

The determination of the spectrum of eigen energies is considered in this paper to be the result of a stochastic process. By repetitive measurements, a probability $P_i$ can be assigned to find the system at a specific eigen energy $E_i$. In an experiment this is achieved by counting the number of (uncorrelated) occurrences of $E_i$. In the discrete case the relative frequency of occurance ($P_i$) of eigen energies $E_i$ determine a discrete spectrum whereas in the continuous case the values of $E_i$ are replaced by a density of states \cite{Bransden}. For a quantum mechanical ensemble in thermal equilibrium the probability $P_i$ is proportional to $D(E_i)$ where $D(E,kT)$ is the proper quantum distribution function of energies of the ensemble \cite{Bransden} under consideration at thermal energy $\beta=kT$ under conditions of thermal equilibrium. The energy distribution $dn(E)/dE$ of the ensemble is defined by the product of the single particle density of states $\rho(E)$ and the quantum distribution function \cite{Reif,Bransden}. For a dense, free fermion ensemble the energy distribution is according to:
   
\begin{eqnarray}
\frac{dn(E)}{dE}=\rho(E)D(E)= \frac{\sqrt{2}m^{3/2}}{\pi^2\hbar^3}\sqrt{E}
\frac{1}{e^{\beta(E-E_f)}+1}
\label{fermi-dirac} 
\end{eqnarray}

where $m$ is the mass of a single particle in the ensemble and $E_f$ the Fermi energy \cite{Reif,Ashcroft,Bransden}. The latter equation is our model for the energy distribution of free electrons $(dn_e(E)/dE)$ for a systems with total charge $Q<0$ or the energy distribution of free holes $(dn_h(E)/dE)$ for a system with $Q>0$. The characteristical shape of the energy distribution allows to make a comparison of experimental data within the theoretical framework of free fermion ensembles. We will use eqn.~\ref{fermi-dirac} to obtain the temperatures of electron and hole gases that are contained near the surface of a charged metal electrode. 

The Fermi level of an ensemble of electrons is determined by the energy of the two last added particles \cite{Reif,Ashcroft}. Since the energy of a single electron is given by $E=-qV$ \cite{Jackson}, where $V$ is the static electric potential and $q=1.602~10^{-19}C$, the Fermi energy of an electron ensemble is given by $E_f=-qV$. Here we make use of the fact that the static electrical potential of an metal object is a measure of the Fermi level of the electron ensemble that is confined at its outer surface. The quantum mechanical response of the Fermi energy upon exchange of a single electron is determined by the density of states of surface charge (quantum capacitance per unit of area) given by $c_{qm}=q^2m/\pi\hbar^2=4\epsilon_o/a_o$ \cite{Ashcroft} where $a_o$ is Bohrs' radius. The change in the electrostatic energy of a macroscopic amount of charge (ensemble) is determined by the classical density of states of surface charge (capacitance per unit of area). For a (macroscopic) capacitor with radius $r$ this is given by $c_{cl}=\epsilon_o/r$ \cite{Jackson}. Thus, the change in energy by the release of a single electron charge (quantum mechanical regime) is identical to the release of a macroscopic amount of charge by an ensemble (classical regime) of magnitude $Q=4r/a_o$.

To obtain an electron energy spectrum rather than an expectation value, we detect fluctuations in the static potential of the metal electrode rather than averages. The electron energy distribution is obtained by repetitive sampling of uncorrelated readings of the static electrode potential. Random fluctuations in the energy of a system at stationary conditions are assumed to obey the Wiener-Khinchine relations \cite{Reif}. The correlation function $K(s)$ of random temporal fluctuations in the energy $E(t)$ of an ensemble is given by the self convolution $K(s)=\left< E(t)E(t+s)\right>$ \cite{Reif}. This correlation function is the Fourier transform of the (power) spectral density of $E(t)$ which is $P_i$ for a discrete quantum system and $dn(E)/dE$ for a continuous system. The spectral density $dn(E)/dE$ can be readily obtained in an experiment using the following procedure: the number of samples $dn$ that is found in the interval $-q(V,V+dV)$ are assigned to the energy distribution function $dn(E)/dE$ in the interval $(E,E+dE)$. In other words: to obtain the electron energy distribution we build a histogram from observed series of fluctuating voltages. 

\section{Experiment}
\label{sec:Experiment}
\begin{figure}[t]
\center
\includegraphics[angle=-90,width=90mm]{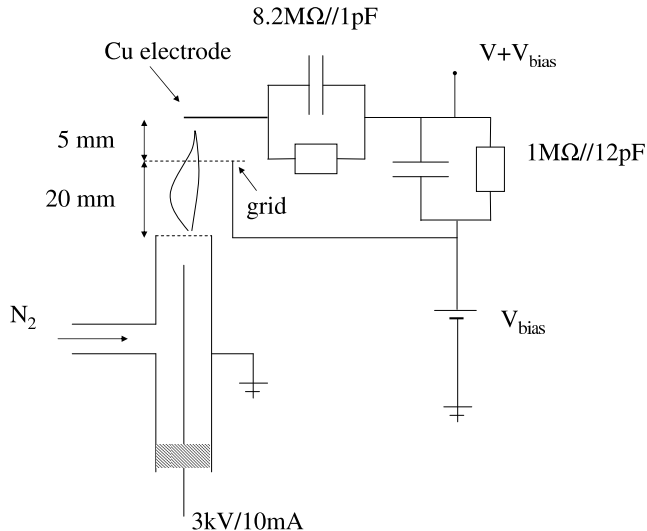}
\caption{\label{set-up} Overview of the experimental set-up. The static electric potential of a Cu electrode is perturbed by the afterglow of a stationary jet of excited state helium or nitrogen produced by a high voltage discharge. The electrode is located 25 mm above the gas outlet. A grid near the Cu electrode, held at a fixed potential, is used to remove space charge surrounding this electrode. The system is operated at ambient conditions. Events that change the potential of the electrode are recorded by a digital oscilloscope as changes in the voltage $V(t)$.}
\end{figure}

The experimental set-up of the detector that is used in this work is depicted and explained in figure~\ref{set-up}. A Cu wire (30x1mm) serving as an electrode is configured as a large bandwidth, high impedance voltage probe (see Appendix). Its electrical potential is perturbed (figure~\ref{fig:data}) either by excited state molecular nitrogen or excited state (atomic) helium produced in a cold plasma operated at ambient conditions. Specific features concerning the afterglow of nitrogen (Vegard-Kaplan system) are discussed in more detail in the Appendix. A preset bias voltage is used to add electrons ($Q>0$) or holes ($Q<0$) to the ensemble held at the surface of the Cu electrode.

\begin{figure*}[t]
\begin{tabular}{cccc}
\includegraphics[width=45mm, angle=0]{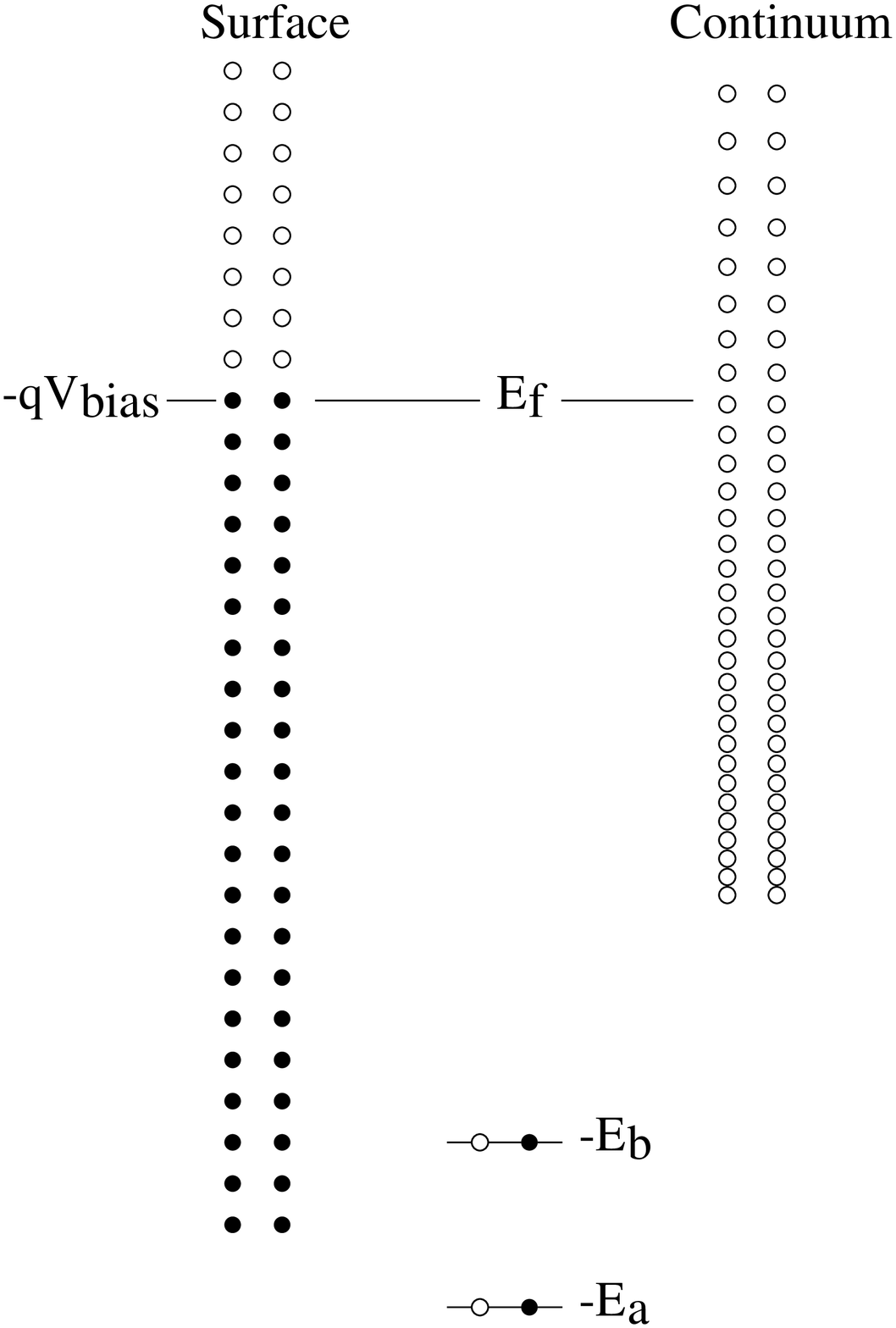}&
\includegraphics[width=45mm, angle=0]{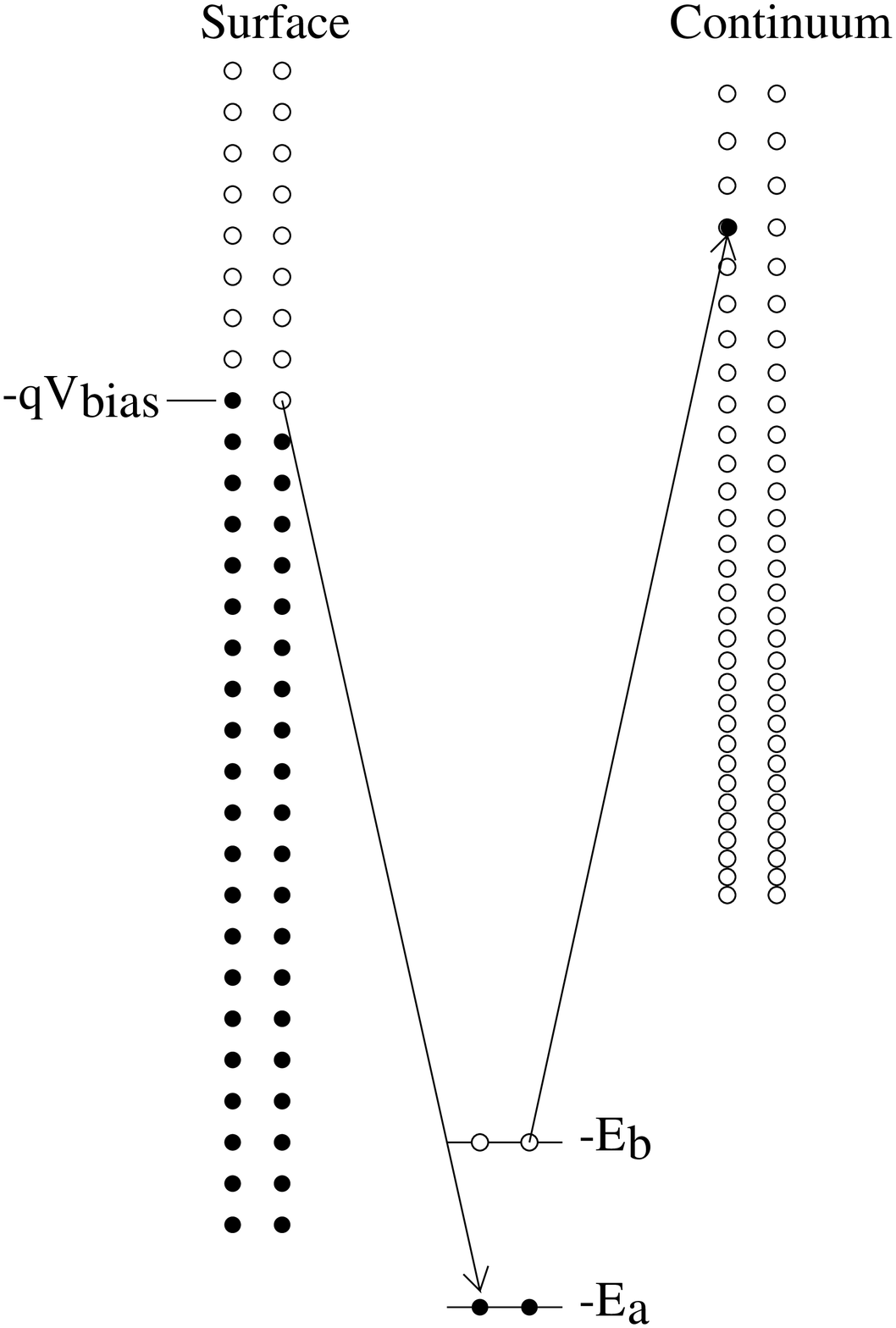}&
\includegraphics[width=45mm, angle=0]{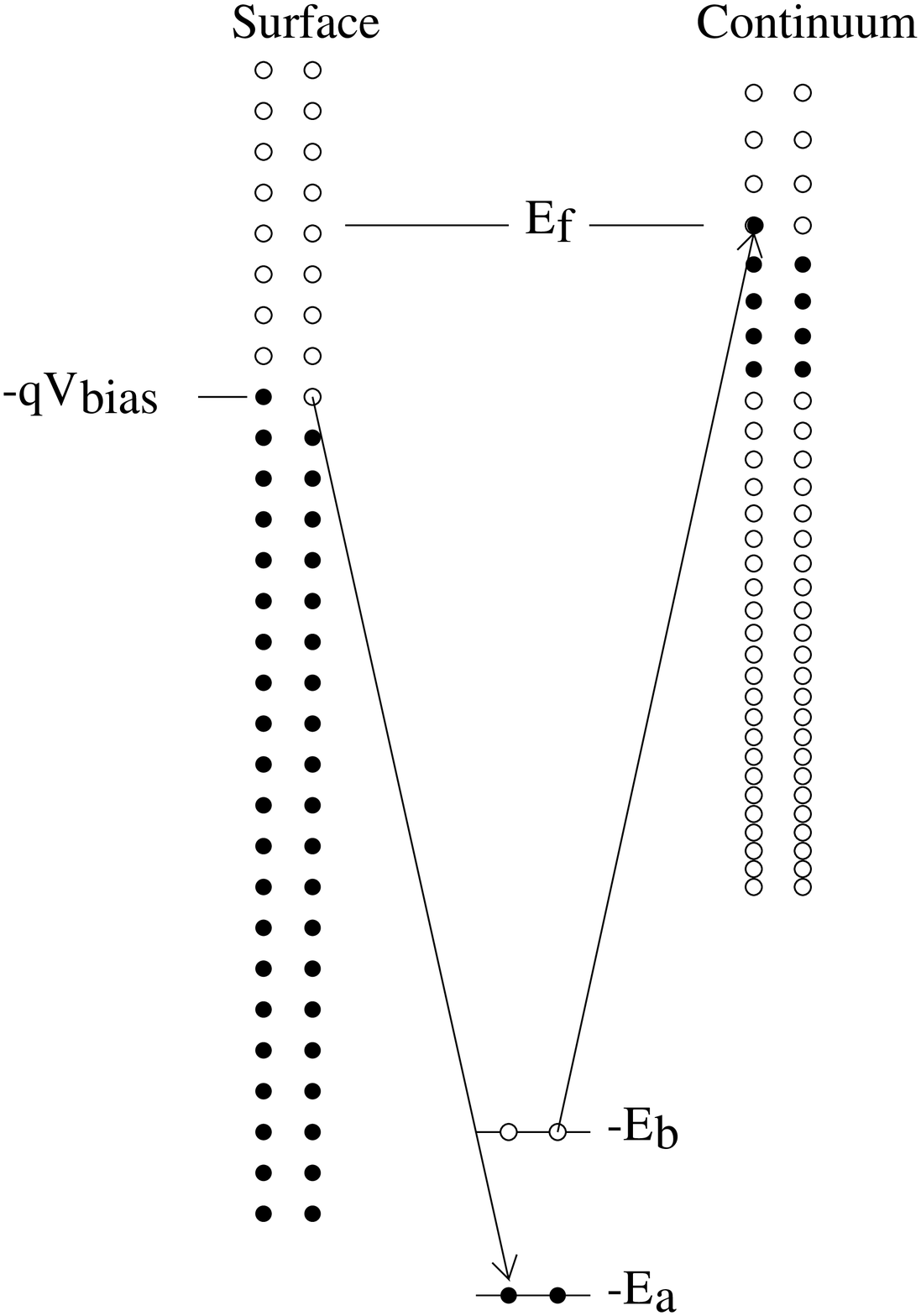}&
\includegraphics[width=45mm, angle=0]{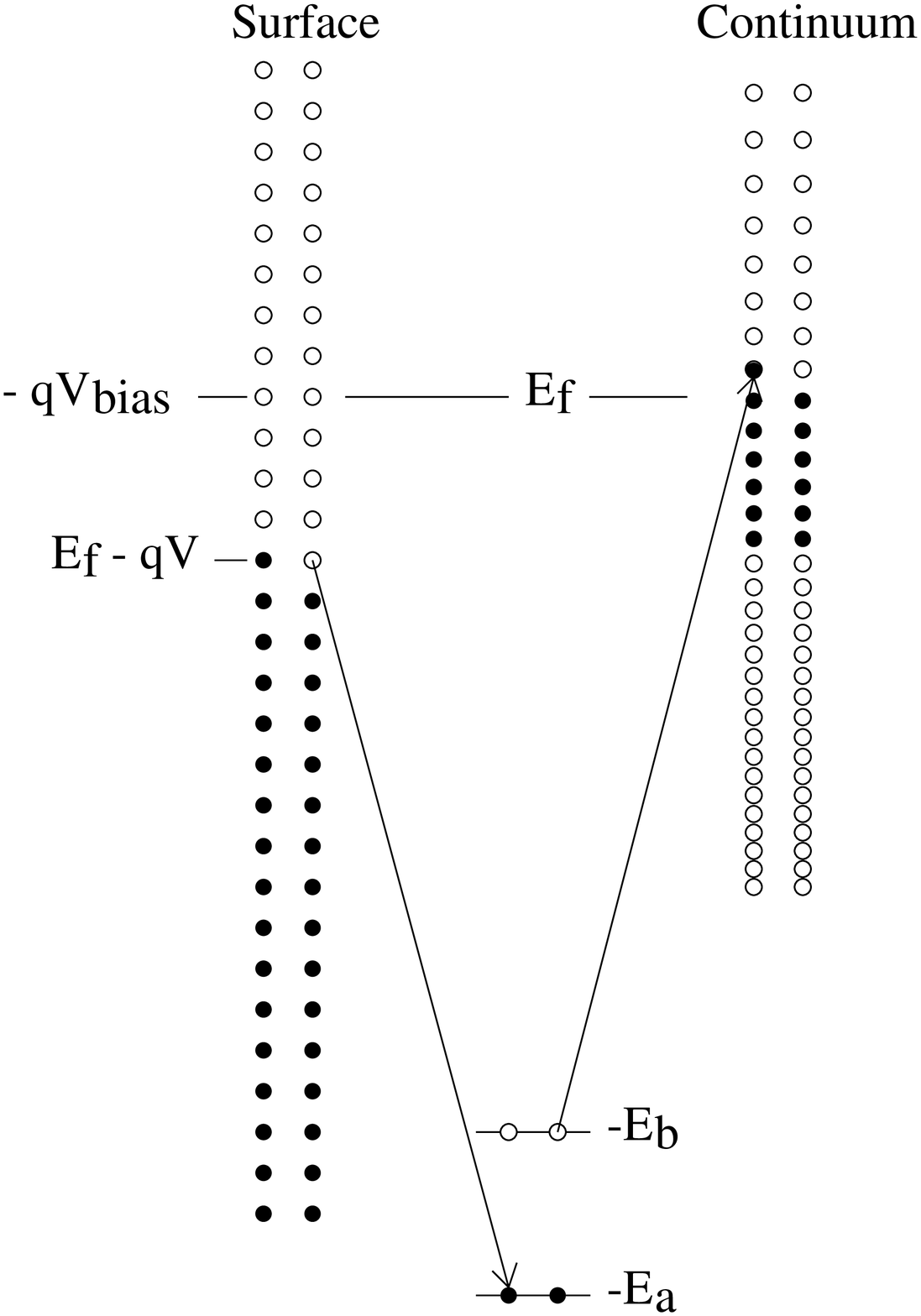}\\
(A)&(B)&(C)&(D)\\
\end{tabular}
\caption{\label{Auger}Schematic of the Auger de-excitation process of an exited state molecule near the surface of a negatively biased metal. Open circles represent hole states and black dots electron states. (A) An excited molecule, represented here as a two level $(E_a, E_b)$ electron system, approaches a metal represented by its surface states. (B) At close range an electron, originally held at the surface, is transferred to the ground state of the molecule triggering the emission of the Auger electron into the continuum \cite{Hotop}. (C) The emission of multiple electrons at ambient conditions leads to the build up of space charge and elevates the Fermi level of the combined system. This process continuous until the Fermi level has increased by an amount of $-qV=E_b-E_a$. (D) If the excess of space charge is removed, the initial Fermi level of the system is maintained throughout the process. An event is then recorded as an increment in the voltage of the detector by an amount $V$.}
\end{figure*}

Redox events at the electrode surface involving the exchange of charge, lead to a temporary change in the static potential of the electrode. The corresponding changes in voltage were recorded. The energy distribution of either electron or hole ensembles were recovered from these recordings by building a histogram as discussed above. Voltages were typically sampled at a rate of 1 kHz. Excessive space charge (figure~\ref{Auger}) that builds up during the experiment was removed using an alumina grid (20x20mm, mesh 3mm, 70\% open). 

The observed events, fluctuations in the voltage, are triggered by the Auger process at the metal-air interface when excited state nitrogen molecules are de-excited by surface electrons upon approach \cite{Hotop} as depicted in figure~\ref{Auger}. Note that by applying this method we directly detect the energy distribution of the electron ensemble near the electrode rather than the energy state of a single electron that is first extracted and then analysed on the basis of its kinetic energy (under vacuum conditions) \cite{Hotop,Bransden}. This feature enables operation at ambient gas conditions. At a positive biased detector a similar process occurs that results in the detection of the energy distribution of positive charges (holes).

\begin{figure*}[t]
\centering
\begin{tabular}{cc}
(A) &\includegraphics[width=85mm, angle=-90]{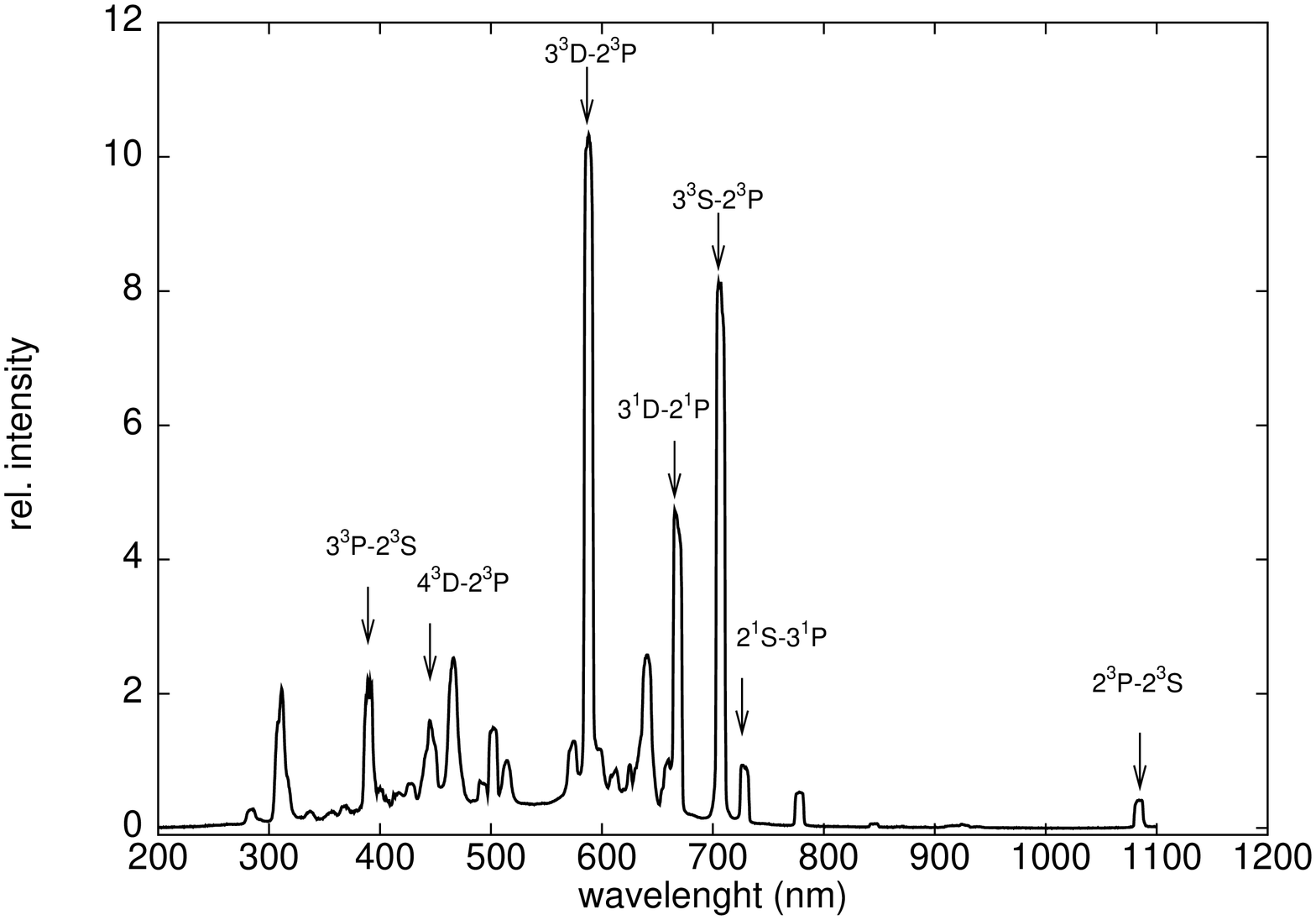}\\
(B)& \includegraphics[width=85mm, angle=-90]{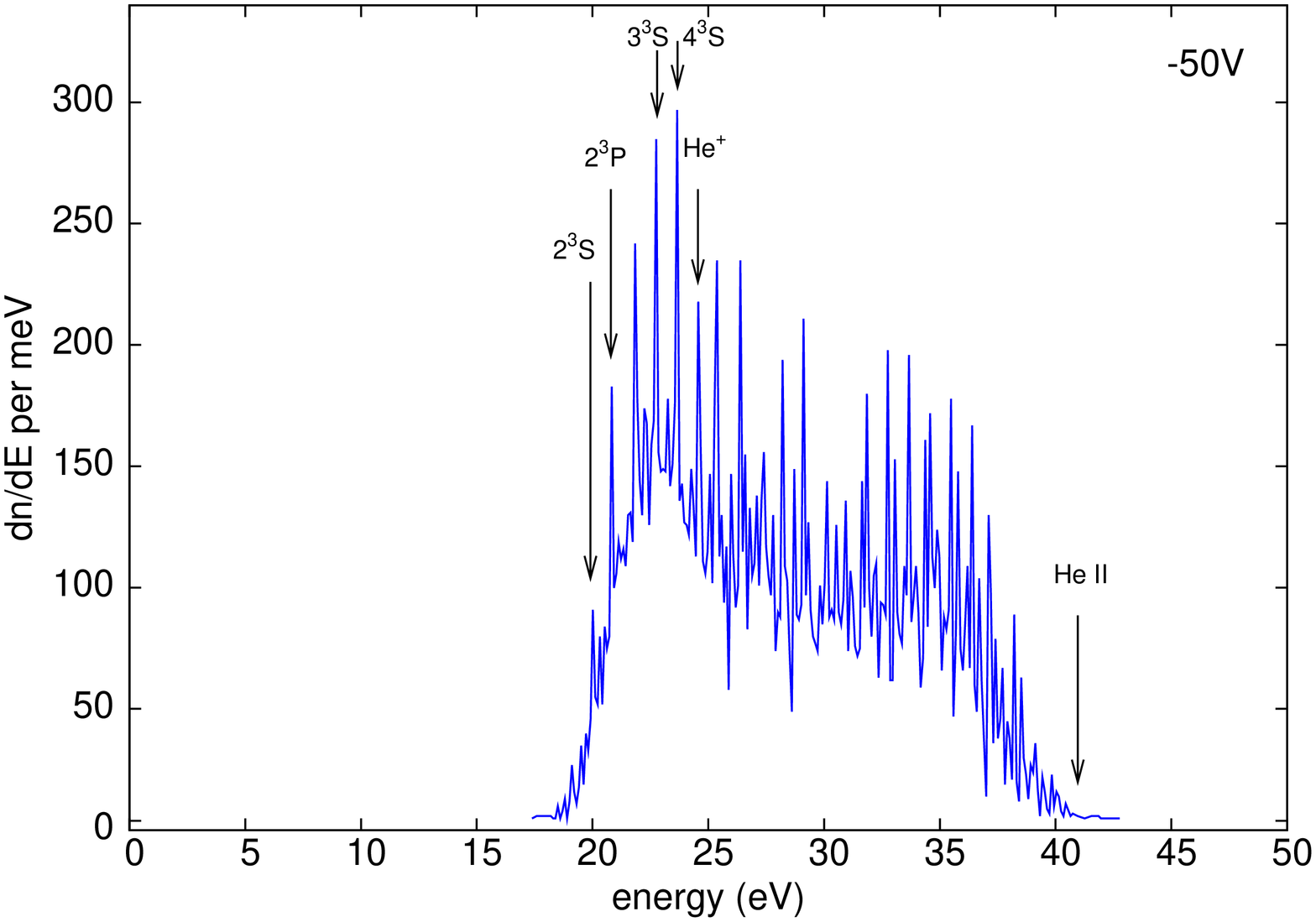}\\
\end{tabular}
\caption{\label{fig:helium}Observed photon- and electron energy spectrum in a helium gas discharge operated at 300K and 1 bar. (A) The primary content of the plasma is singly excited triplet helium as shown by the light emission spectrum. (B) The electron energy distribution that was obtained after analysis of the fluctuations in the static potential of the Cu electrode. As the valence electron in excited state helium can be considered as a free electron, the measured electron energy distribution coincides with the energy level diagram (density of states) of He~I \cite{Nist}. Note that no electron states were detected in the range of 0-19.8eV. In the range of 19.8 to 24.6eV the lowest lying triplet states of He~I were resolved and ascribed as indicated. The structure in the observed continuum states, ranging from the ionisation potential of He~I at 24.6eV up to the threshold energy of He~II at 40.8eV, is attributed to molecular states of trace gases other than helium.}
\end{figure*}

The assumed relation between the single particle electron energy and the changes in the static potential of the detector, as discussed in section~\ref{sec:Theory}, requires experimental proofing. We have validated that the observed random fluctuations in the static potential directly correspond to the electron density of states using the He~I system. This was achieved by replacing the carrier gas N$_2$ by He (Hoekloos-Linde, gas purity grade 5.0). The primary content of the helium plasma that was obtained is singly excited, triplet helium as indicated by the light emission spectrum given in figure~\ref{fig:helium}A. This was confirmed by the measurement of the electron density of states depicted in figure~\ref{fig:helium}B. The electron energy distribution was obtained at a bias of -50V by analysis of the fluctuations in the electrostatic potential of the Cu electrode similar as observed in the case of nitrogen and similar as discussed in section~\ref{sec:Theory}. The electron density states of the free electron gas is a direct measure of the energy level diagram of He~I \cite{Nist}. The energy resolution of the spectrum is 81meV (bin size of the histogram), sufficient to resolve the lowest lying triplet states in the range of 19.8-24.6eV as indicated. The structure that was observed in the continuum states of He~I, as valence electrons associated to He$^+$ ranging between 24.6 and 40.8eV, is attributed to the molecular states of contaminating gases present in trace amounts. Note that we have an exact match of the electron energy distribution (a histogram of the raw data) with the known electron energy levels of He~I, i.e. the observed spectrum is not corrected for an offset by the work function as in conventional surface Auger spectroscopy \cite{Hotop}. It is remarkable having observed the electronic eigen energies of helium in the simplified surface Auger experiment under evaluation. Apparently, the stationary fluctuations in the surface potential of a metal are directly connected to the eigen energies of the quantum system directly near the outer surface of the metal. In a first attempt to understand the thermodynamical aspects of this phenomenon, we have studied the effect under consideration in more detail using molecules (N$_2$) instead. Unlike atoms, electrons in molecules can be treated as harmonic oscillators. Since the electronic energy levels in molecules are evenly and closely spaced (typically 200meV), the electron density of states can be approximated by a continuous distribution i.e. we expect that the valence electrons of molecules behave as an idealized free electron gas in thermal equilibrium as suggested by eqn. ~\ref{fermi-dirac}. 

\section{Observations and analysis}
\label{sec:Observations and analysis}
\begin{figure*}[t]
\centering
\begin{tabular}{cc}
\includegraphics[width=95mm]{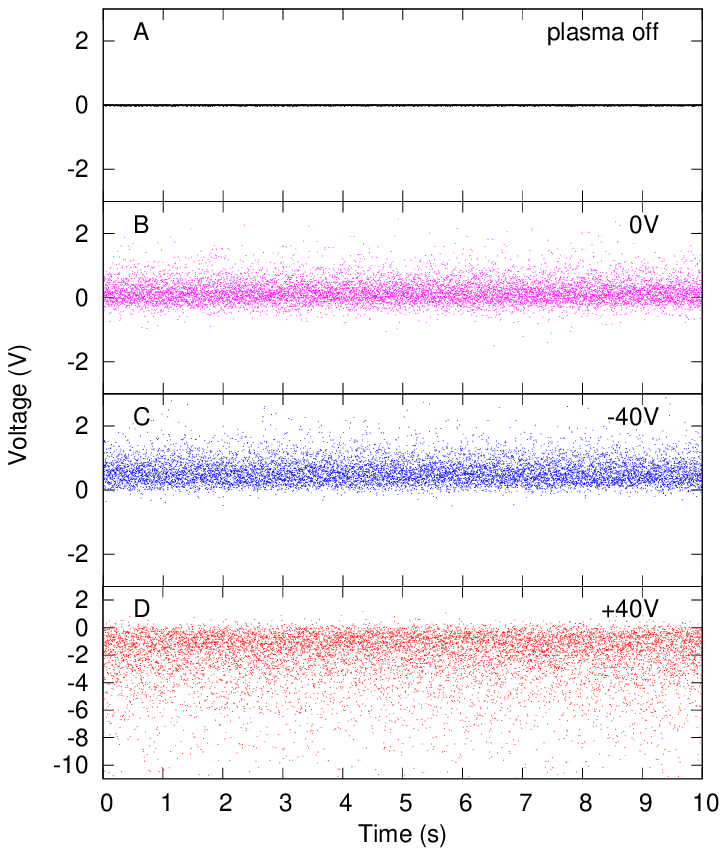}&
\includegraphics[width=95mm]{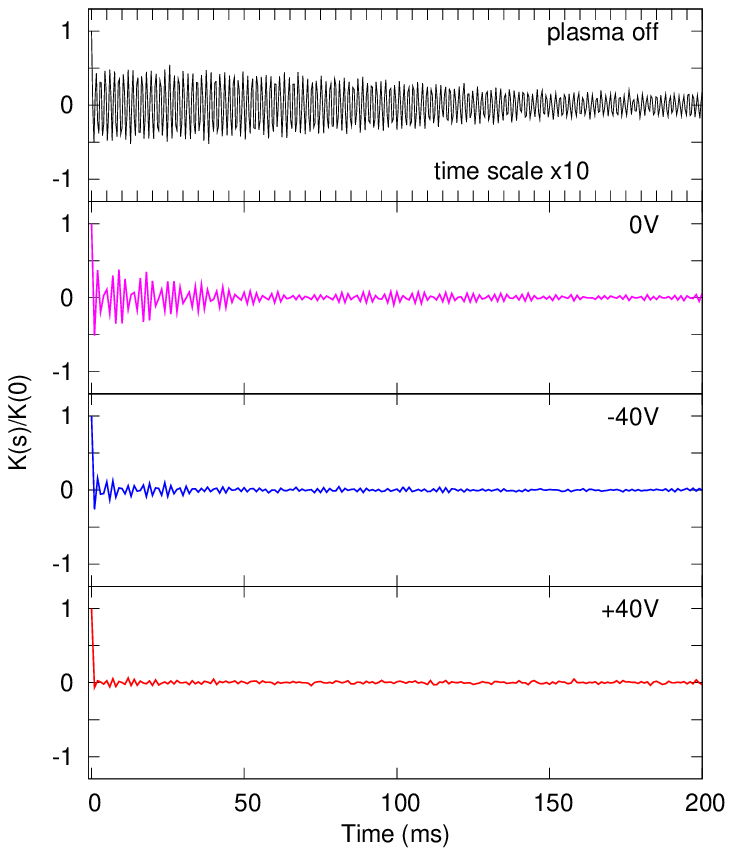}\\
\end{tabular}
\caption{\label{fig:data}Typical fluctuations (left) and autocorrelations (right) that were observed in the electrical potential of the Cu electrode exposed to a nitrogen plasma. (A) Base line which is recorded when the plasma source is turned off (B) Observed fluctuations when the plasma is turned on for the case of a zero biased detector, (C) a negative biased and (D) a positive biased detector. The autocorrelation time decreases with increasing fluctuations in the voltage. When the plasma source is activated the correlation time is reduced to less than 10 ms. For a biased detector the correlation time is typically less than the time between successive measurements. Under these conditions voltage readings are uncorrelated and readings can be considered as independent outcomes of a stochastic process.}
\end{figure*}

In figure~\ref{fig:data} we present a typical series of observed fluctuations in the electrode potential at fixed bias voltage when exposed to a nitrogen plasma. The stationary fluctuations that were studied were invoked only when the electrode was held in the region of the plasma jet were light emission occurs and completely disappeared when the electrode was held just outside this region. Next, the amplitude of the observed fluctuations increase with the electrode potential. The possible role of unintended electronic interference could be ruled out by switching the gas discharge ´off´ while observing the afterglow of the nitrogen plasma. The fluctuations could be observed during the decay of metastable nitrogen (see Appendix). Therefore, we can conclude that the detected fluctuations are no artifacts.

\begin{figure*}[t]
\centering
\includegraphics[width=140mm]{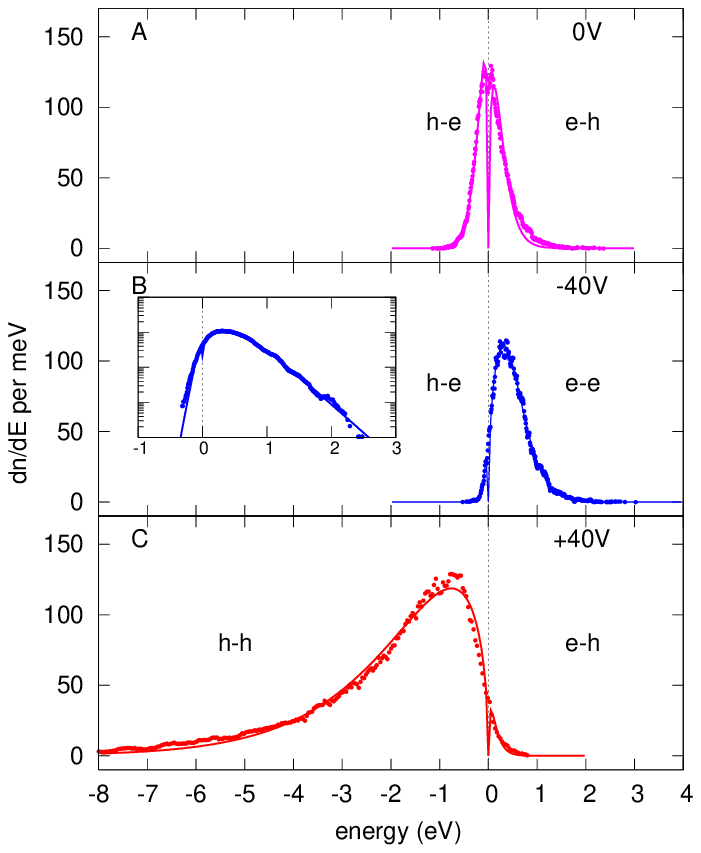}
\caption{\label{eandhgas} Energy distributions derived from the observed fluctuations in the static potential of a nitrogen plasma. The energy distributions (data points) are fitted (solid lines) according to eqn.~\ref{fermi-dirac}. (A) The energy distribution at zero bias is the resultant of two components which were separated by shifting the Fermi level of the system. (B) At negative bias electrons are added to the system. The energy distribution is observed is a free electron gas (e-e) for $E>0$ and a minor tail for $E<0$ which is due to the recombination process of an hole in an electron density (h-e). The inset at a log scale demonstrates the proper asymptotic behavior of the model. (C) At positive bias holes are added to the system and a free hole gas (h-h) is observed for $E<0$ and a minor tail for $E>0$ which is the recombination process of a electron in a hole density (e-h). In this paper we suggest that the energy distribution that is observed at zero bias is the result of two processes: for $E<0$ it is an hole decaying in an electron density and for $E>0$ it is an electron decaying in a hole density.} 
\end{figure*}

The observed fluctuations at negative bias (figure~\ref{fig:data}C) relate to the emission of charge such that the voltage of the detector is increased. Under these conditions, electrons that were initially located at the surface of the electrode are transferred to the continuum by the Auger process (figure~\ref{Auger}). These events result in a temporary decrease in the Fermi level and correspondingly to a temporary decrease in the voltage as explained by the process discussed in figure~\ref{Auger}. A similar process occurs at positive bias. Electron energy distributions were obtained after building a histogram of the observed fluctuations. Typical results for zero, negative and positive bias are presented in figure~\ref{eandhgas}. The energy resolution, defined by bin size of the histogram when sorting the recorded voltages, was 200meV.

The autocorrelation of stationary fluctuations was found to decrease with increasing fluctuations in the voltage. The correlation time of the signal when the plasma was turned 'off' (figure~\ref{fig:data}A) exceeds 1 second. This was due to the finite resolution of the recording digital oscilloscope. The data in figure~\ref{fig:data} were acquired at a fixed setting of sensitivity of the oscilloscope. The detection resolution under these conditions was 20mV. The residual fluctuations in the voltage of figure~\ref{fig:data}A (thermal noise) when measured at 0.1mV resolution yields 23$\pm$2mV (FWHM) (data not shown). Since the resolution of the measurements was nearly equal to the thermal noise level, the majority of successive readings of the voltage in figure~\ref{fig:data}A were assigned the same value in the analog-to-digital conversion. This resulted in the extended correlation time as observed in figure figure~\ref{fig:data}A. Part of this effect, which is most pronounced at small amplitudes, is transferred to the fluctuations observed in figure~\ref{fig:data}B. For signals with a large signal-to-noise ratio this effect is eliminated and an autocorrelation time is obtained that is typically less than the time between to successive readings. The correlation data in figures~\ref{fig:data}C and~\ref{fig:data}D satisfy $K(t)/K(0)<1/2\pi$ for $t>0$ i.e. the readings in these series are uncorrelated.

\subsection{Coarse structure of the energy distribution}
We have recorded and analysed voltage fluctuations for a series of bias voltages in the range of -40 to +40V. The observed energy distributions of the two fractions were fitted to the energy distribution given by eqn.~\ref{fermi-dirac} to obtain the temperatures of both electron and hole ensembles. At negative bias voltages (excess of electrons) we associate the energy distribution with a free electron gas. This is illustrated in more detail in figure~\ref{eandhgas}B. We distinguish the elastic electron-electron scattering process (e-e) that occurs for $Q<0$ and $E>0$ from the inelastic hole-electron recombination process (h-e) that occurs for $E<0$.  Likewise, for a positive bias (excess of holes) we associate the energy distribution to a free hole gas (h-h) for $Q>0$ and $E<0$ and the decaying tail for $E>0$ that is the result of electron-hole (e-h) interactions (figure~\ref{eandhgas}C). The energy distribution at zero bias is the composite of a residual free hole gas and a residual free electron gas. We suggest that the observed energy fluctuations at zero bias ($Q=0$) is the direct result of two recombination processes: for $E>0$ we observe the energy distribution of holes decaying in a free hole electron density (e-h) and for $E<0$ we observe the energy distribution of an electron ensemble decaying in an free hole density (h-e). In order to investigate these assumptions we have made an estimate of the lifetime of electrons and holes. We associate the energy distribution of an ensemble to the single particle lifetime using the Laplace transform. Under these assumptions the width of the observed energy distribution ($\Delta E$) is a measure of the rate constant \cite{Reif} of the recombination process according to 

\begin{equation}
\Delta E=\hbar\Gamma=\frac{\hbar}{\tau}=kT.
\label{Laplace}
\end{equation}

The recombination lifetime of electrons and holes under the distinct conditions discussed in this work are estimated in table~\ref{table1}. The lifetime of an electron at charge neutral conditions (zero bias) is 10.8$\pm$0.5 fs and lifetime of a hole 4.4$\pm$0.2 fs. The lifetime of a hole in an electron density yields 14$\pm$2 fs whereas the lifetime of an electron in an hole density yields 6.3$\pm$0.2 fs.    

The behavior of the temperature of the electron gas and the hole gas was analyzed in more detail (figure~\ref{tempbias}). The temperatures of the hole gas and electron gas were determined (see Appendix) for two series at 15 different bias voltages ranging from -30 to +50V. The two data series shown in figure~\ref{tempbias} are the average of two independent measurements: the temperature at each bias voltage is a two point weighted average. The ensemble temperature of the hole gas at negative bias is nearly a constant at the level $kT$=47$\pm$6 meV. At positive bias the temperature inclines as a function of the bias voltage. The temperature of the electron gas is nearly a constant, fixed to a level $kT$=159$\pm$4 meV at negative bias. The explanation of latter behavior originates from the fact that the temperature of a degenerate gas is independent from the Fermi level. Actually, the chemical potential is independent of the temperature \cite{Reif,Ashcroft}. This is different for a gas at low densities (degenerate case). Upon raising the Fermi level under these conditions the kinetic energy of particles is continuously redistributed by elastic collisions within a range of $\Delta \sim kT$, thus leading to an increase in the temperature when particles are added. For a single component gas the temperature dependence of an degenerate gas was estimated by the elastic scattering rate of binary collisions \cite{Reif}:

\begin{equation}
\Gamma=1/\tau=\frac{1}{2} \rho_o \sigma v.
\label{elastic}
\end{equation}

where $\rho_o$ is the density of scatterers, $\sigma$ is the cross-section of the process and $v$ the relative velocity of the particles. The factor of $\frac{1}{2}$ is to account for the symmetry of the collision system (e-e or h-h). The relative velocity of two particles is given by $v=\sqrt{2E_f/m}$. The density multiplied by the cross section in eqn.~\ref{elastic} can be expressed in terms of $\sqrt{E_f}$ (see eqn.~\ref{scattering rate}). At high temperature (low density) conditions the width of the energy distribution of identical particles is then given by $\hbar\Gamma=kT=0.391 E_f$. Note, that this result is nearly identical to the thermal energy per particle given by $kT=\frac{2}{5}E_f$ in case of a free degenerate fermion gas \cite{Ashcroft}. In our experiments the temperature of the free hole gas is assumed to present the equilibrium of a cold fraction (hole states in the ground state of $N_2$) with a thermal energy $kT$=25meV ($T$=300K) and a hot fraction (hole states in $N_2^*$) at thermal energy $kT=0.391 E_f$. In case of a diluted hot fraction, the equilibrium temperature of the mixture is a linear function in the number of particles present in the hot fraction (ideal gas). In this work we have fitted the temperature dependence of the hole data in figure~\ref{tempbias} to a linear function in the Fermi level to obtain a proportionality factor (experimental value) according to $kT=(0.019\pm0.001)E_f$. 

Typical free electron and free hole densities that occur in a nitrogen plasma were estimated using eqn.~\ref{elastic} at the corresponding temperatures listed in table~\ref{table1} under conditions of zero bias. The electron density in the experiments evaluated for $kT_e$=149$\pm$8~meV yield $\rho_e=2.6\pm0.2~10^{22}$~cm$^{-3}$ and the hole density $\rho_h=0.68\pm0.05~10^{22}$~cm$^{-3}$ for $kT_h$=61$\pm$3~meV.

\begin{figure}
\centering
\includegraphics[width=90mm]{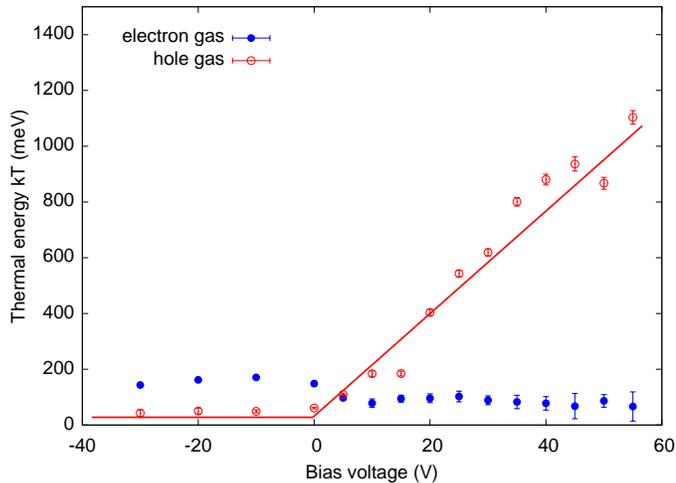}
\caption{\label{tempbias} Temperatures of the observed free electron and free hole ensembles in a nitrogen plasma as a function of the bias voltage. The model for the h-h process is (solid line) is a straightforward extension of eqn.~\ref{elastic} and relates to an ideal free fermion gas at high temperatures.}
\end{figure}

\subsection{Molecular structure}
\label{sec:Molecular structure}
At high energy resolution, sharp, peaked structures were observed that occur at specific energy levels similar to the case of helium. The data shown in figure~\ref{structure} is a typical energy distribution of a hole gas for a nitrogen plasma detected at a resolution of 25 meV. The peaked structure in figure~\ref{structure} resembles a molecular vibrational structure. If we account for the specific experimental conditions that were imposed, we deduce that these structures probably originate from redox reactions of exited state molecular nitrogen at the charged Cu electrode. This structure was also observed in the energy distribution of the electron gas. The observed free electron states coincide with the vibrational states of $N_2(X^1\Sigma_g^+,\nu=0,..,5)$ well below the threshold energy of $N_2^-$ at 1.6eV.    

\section{Discussion}
\label{sec:Discussion}
\subsection{Electron emission process}
\label{sec:Electron emission process}
In presence of a metal surface the decay of excited state nitrogen is through the process of surface Auger effect. We have carefully checked whether the observed fluctuations as shown figure~\ref{fig:data} were actually due to the activated nitrogen at the electrode under investigation as a result of the plasma jet. More specifically, we have canceled the option that the emissions are the result of field emissions at high electric field strengths (corona discharges). In absence of the plasma, field emissions could only be observed after increasing the electric field strength at the electrode surface to a level in excess of 30 kV/cm. This was achieved after switching to a bias voltage of +(-)15kV using EMCO DX250(N) high voltage power supplies. The observed energy distribution according to eqn.~\ref{fermi-dirac} is certainly different from electrons that are the result from high field field emissions \cite{Ashcroft,Heer,Yang,Bonard}. The energy distribution in field emission follows the Fowler-Nordheim law \cite{Ashcroft} as a result of quantum tunneling through the cathode-anode barrier. Secondly, emissions that have been observed in our work occur in the energy range of 0 to 20eV, which is one order of magnitude less than the threshold reported in \cite{Heer,Yang,Bonard}.

\subsection{Fluctuations}
\label{sec:Fluctuations}
Random fluctuations in energy and the relaxation of an excited state system to its ground state are subject to the fundamental thermodynamic aspects of finding the system in any of the accessible microscopic states \cite{Evans}. Under stationary conditions, energy fluctuations obey the Wiener-Khinchine relations \cite{Reif}. The correlation $K(s)$ of random temporal fluctuations in the energy $E(t)$ of an ensemble is given by the self convolution $K(s)=\left< E(t)E(t+s)\right>$ \cite{Reif}. This correlation function is the Fourier transform of the spectral density of $E(t)$ which is $dn(E)/dE$, if we assume that the electronic energy of a quantum system satisfies $E=\hbar\omega$. The exact relationship is given by

\begin{equation}
\frac{dn(E)}{dE}={\bf\it F} \left[K(s)\right].
\label{Wiener-Khinchine}
\end{equation}

The Wiener-Kinchine relations imply that the energy spectrum of an ensemble of electrons is directly obtained after taking the inverse Fourier transform of the autocorrelation of the time series $qV(t)$. A constraint is that the voltage V(t) has to be recorded using a detector with a bandwidth exceeding $qV/\hbar$. For the experiments under consideration this yields a typical bandwidth of 10 PHz=10$^{16}$ Hz. Such devices do not exist. Therefore, a direct Fourier transformation using $E(t)$ to obtain the density of states is only relevant from an academic point of view. However, employing a detector of such a bandwidth is not a requisite to determine the spectral density. 

A quantum system oscillates between its eigenstates (eqn.~\ref{wavefunction}) at frequencies on the order of $\omega_i=E_i/\hbar$ until the event of a measurement (e.g. sampling the voltage of an electric system) when its eigenvalue $E_i=-qV_i$ is read. During the period that a reading is taken, the system under investigation is in a definite eigenstate which is maintained (in our case by a sample and hold circuit) until a successive reading is performed. Only after performing a successive measurement, the eigenstate will have changed randomly. This random change can be understood by analysing the time dependent response of a detector for signals that change at frequencies larger than the bandwidth. The time dependent voltage can be decomposed into its harmonics according to $V(t)=V_1sin(\omega_1 t)+V_2sin(\omega_2 t)+...$. Samples of $V(t)$ that are recorded at a rate $\omega_s<<\omega_i$ yields the collection of readings of $V(t)$ at random phases. Under these conditions the likelyhood of an observation at level $V$ is given by the probability density function of $V(t)$ yielding $D(V)=1/(\pi\sqrt{V_i^2-V^2})$. Note that $D(V)$ is just a limiting case of the Dirac delta distribution $\delta(V_i-V)$. Thus, the outcome of a series of measurements are static, single valued (quantised) readings. The expectation value for the energy of a free electron in an eigenstate is given by $\left<E_i(t)\right>=\left<\psi_i(0)\right|H\left|\psi_i(t)\right>=\left<E_isin(E_it/\hbar)\right>$. Using $E_i=qV_i$ we obtain $\omega_i=E_i/\hbar=qV_i/\hbar$. Thus, a free electron with energy $E_i$ gives rise to a harmonic signal of frequency $\omega_i=qV_i/\hbar$ and amplitude $V_i=E_i/q$. As in the present work $\omega_i$ is much larger than the bandwidth of the detector, the measurement yields the recording of a static value: $\left<E_i(t)\right>=\delta(E_i-E)$. In conclusion: the detection of the eigen energies of electrons yields the random observation of the amplitudes of high frequent harmonics, manifested as fluctuations in the static potential.

This means that a spectrum of electronic eigen energies can be obtained by the method employed in the manuscript, consistent with eqn.~\ref{Wiener-Khinchine}. The only requirement is that the response time of the voltage probe is short compared to the time between successive measurements. In the experiments that have been presented this constraint was met since the response time of our probe which is RC=1pF$\cdot$8.2M$\Omega$=8.2$\mu$s (figure~\ref{set-up}), the time needed to charge the probe, is much less than the time (=1 ms) between successive measurements.

The probability that a transition occurs from a state with its energy in the interval $(E,E+dE)$ to the ground state, is proportional to the density of states at energy $E$ by virtue of Fermi's Golden rule \cite{Reif,Bransden,Feynman}. It is therefore sufficient to determine the relative frequencies of uncorrelated readings of $E(t)$ in the interval $(E,E+dE)$ for a large number of measurements. The energy distribution $dn(E)/dE$ is readily obtained by building a histogram of a sequence of readings of $E(t)$. When voltage readings are uncorrelated (figure~\ref{fig:data}) the density of states of a system is obtained by sampling and counting random values of the energy as explored in this paper. Note that a complete reconstruction of the electronic wave function, a procedure of paramount interest to quantum computing \cite{Bayar}, cannot be performed exclusively on the basis of known energy eigenstates. A complete reconstruction of the wave function requires, in addition, the determination of the relative phases between the different eigenstates. These phases could not be determined as a result of the limited sample rate (bandwidth) of the detection system in present work.

\begin{figure*}[t]
\centering
\includegraphics[width=130mm, angle=-90]{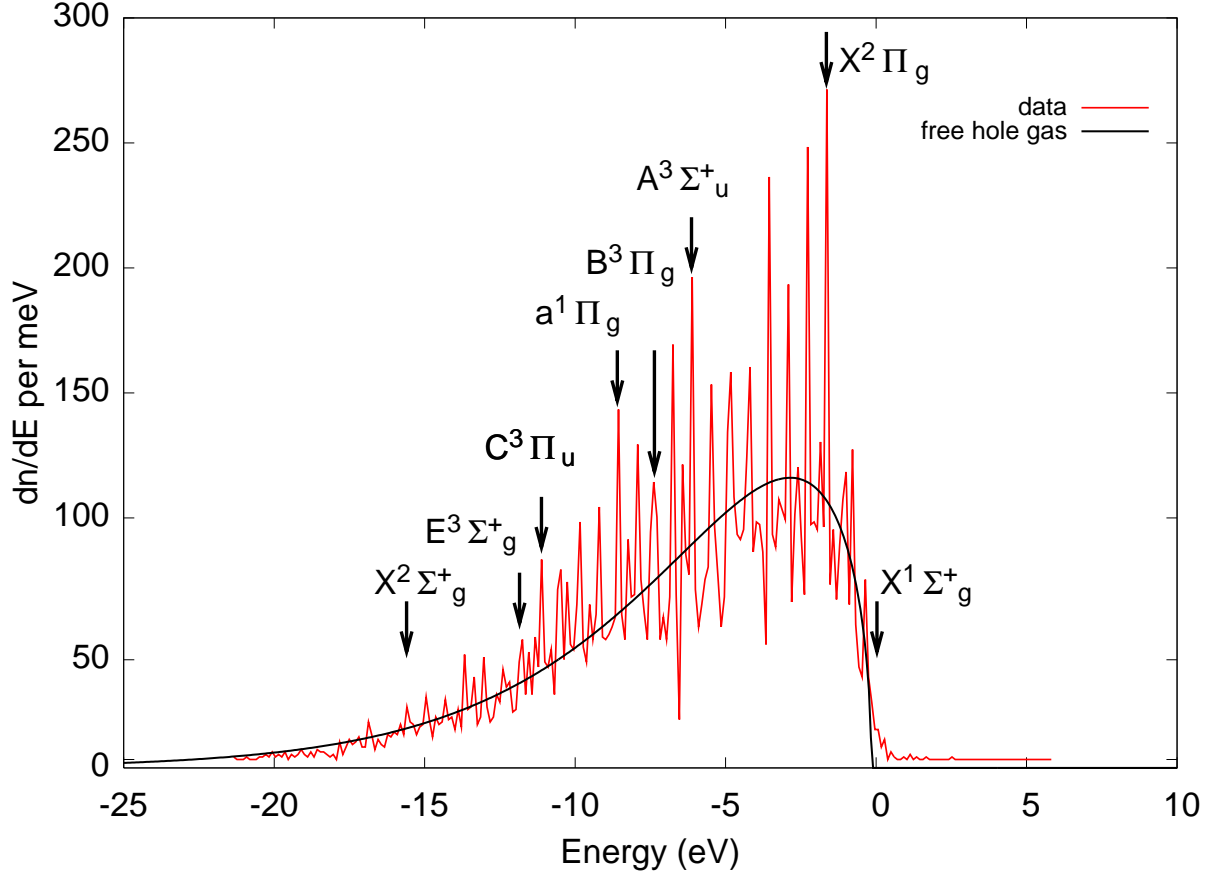}
\caption{\label{structure} High resolution plot of the energy distribution of a free hole gas observed in a nitrogen plasma, revealing residual structure. The peaked structure coincides with known reduction states of excited state nitrogen that originates from the surface Auger process at the metal target. The molecular symmetry and electronic energies of the lowest lying $N_2^*(\nu=0)$ states have been indicated. The ground state level of the molecular ion $N_2^+$ lies at 15.6eV.}
\end{figure*}

The peaked structures that were observed in all of our experiments was analysed in more detail using the typical distribution of hole energies shown in figure~\ref{structure}. As it was confirmed by lifetime measurements and photospectroscopy that the nitrogen jet primarily consist of excited state molecular nitrogen, the electron spectrum in figure~\ref{structure} should contain part of the Auger spectrum of exited state molecular nitrogen at the Cu electrode surface \cite{Hotop}. In figure~\ref{structure} we have marked the onset energies (N$_2^*(\nu=0)$) of characteristic vibrational states of the different molecular symmetries indicated \cite{Lofthus}. We conclude that the energy distribution of the observed free hole gas is partly associated to the reduction states of molecular nitrogen.

\subsection{Interaction of electrons and holes}
\label{sec:Free Hole gas}
Our aim in present work was the decomposition of a single redox reaction into its two fundamental half reactions. One reaction involving the addition of electrons (reduction of the free hole gas) and a second reaction involving the addition of holes (oxidation of the free electron gas). It is well known that an oxidation and its corresponding reduction process at thermal equilibrium can proceed at different energies. An example is photosynthesis by micro-organisms \cite{Slegel} where the energy difference between two different redox states is balanced by the photon energy. The fundamental interaction process of an electron and a hole is recombination \cite{Ashcroft,Feynman}. In the case of a vacuum, recombination is proceeds by annihilation where an electron-hole pair recombines under emission of two 511keV photons (conservation of energy). Under the experimental conditions in this work, recombination of free electrons and holes that are associated to exited state metastable nitrogen occurs through spontaneous decay by a singlet-triplet transition. The (kinetic) energy and angular momentum in the process of spontaneous decay is conserved by emission of 3eV photons (see Appendix) whereas the rest mass is balanced by the molecule, acting as a third body itself. In the presence of a surface, recombination may occur through the surface Auger process were the vibrational excitation is lost by interaction with the electrode (figure~\ref{Auger}). The prediction of the lifetime of a hole(electron) in an electron(hole) density $\rho_o$ for the latter process is estimated using binary collision theory \cite{Reif} according to 

\begin{equation}
\Gamma=1/\tau=\rho_o \sigma v 
\label{scattering rate}
\end{equation}

with $\sigma=\pi r^2=\pi(4\pi\rho_o /3)^{-2/3}$ the cross-section of the scattering process, $v$ the relative velocity of the electron with respect to the hole given by $v^2=3kT/m$ (m=4.5 $10^{-31}$~kg is the reduced mass of the electron-hole system) and $\rho_o$ the density of scatterers. The density of electrons and holes is estimated by the free electron (hole) density \cite{Ashcroft} which yields $\rho_o=Z \rho_m/M m_p$ with Z=3 (thus accounting for a trivalent $N-N$ bond), M=28 and $m_p=1.68~10^{-27}$kg and $\rho_m=1.25$ kgm$^{-3}$ for a nitrogen environment at 1 bar and 300K. The calculated lifetimes according to eqn.~\ref{scattering rate}, evaluated at the temperatures in the experiments, are given in table~\ref{table1}. These theoretical predictions for the recombination lifetime of holes and electrons are in fair agreement with the experimental values if we take into account the crude estimates that have been made in adapting a free electron model. We conclude that the interactions of the observed free electron and free hole gases may proceed through recombination and can be understood quantitatively.
 
\begin{table}[t]
\centering
\caption{\label{table1} Recombination lifetimes of free electrons and free holes in a nitrogen plasma under the experimental conditions that were distinghuised in this work. Experimental values for the lifetimes are according to eqn.~\ref{Laplace} using the experimental ensemble temperatures. Theoretical estimates of the lifetimes are according to eqn.~\ref{scattering rate}. Experimental values for the lifetimes are according to eqn.~\ref{Laplace} using the experimental ensemble temperatures. Theoretical estimates of the lifetimes are according to eqn.~\ref{scattering rate}.}
\begin{tabular}{cccc}
\hline
Experimental  &	Ensemble temp.& \multicolumn{2}{c}{Lifetime (fs)}\\
conditions    & $kT$ (meV)      & {experiment}&{calculation}\\
\hline
negative bias  &  47 $\pm$ 6  & {14   $\pm$ 2   (h-e)} &{9}\\
zero bias      &	61 $\pm$ 3  & {10.8 $\pm$ 0.5 (h-e)} &{7.5}\\
               & 149 $\pm$ 8  & { 4.4 $\pm$ 0.2 (e-h)} &{4.8}\\
positive bias  & 105 $\pm$ 4  & { 6.3 $\pm$ 0.2 (e-h)} &{5.8}\\
\hline
\end{tabular}
\end{table}
Based on the previous analysis we explore the suggestion that electrons and holes are different entities i.e. at thermal equilibrium hole states are not equivalent to electron vacancies. Dirac's equations for the free electron incorporates the electromagnetic field (photons) and has four solutions. Except for two different spin states two types of solutions are known. One of the solution represents the free electron in presence of the electromagnetic field. The second solution represents the conjugate state which has no physical reality. However, it is generally recognized that this conjugate state (with negative energy solutions) are of significance both physically and mathematically \cite{Feynman}. This fundamental aspect on the behavior of electron and hole ensembles is treated by the theory of Quantum Electro Dynamics (QED). Despite this fact, it is highly controversial \cite{Weinberg} to assert that free hole gases may represent this conjugate state since it is commonly assumed that a direct observation of these states is not possible \cite{Feynman,Weinberg}. The conclusion that is justified, is the distinction between the energy distributions of free particles with negative charge and positive charge. Hence, we have to assign different lifetimes to the processes of elastic scaterring and recombination of electrons and holes. This anomalous behavior of particles of opposite charge is remarkable since replacing $q$ by -$q$ (charge conjugation) in the experiment under consideration should not have affected the scattering rates as observed. 

We observed that the energy distribution of the observed free hole gas may be partly associated to the reduction states of molecular nitrogen. However, the energy distribution can not exclusively originate from the surrounding nitrogen. If this were the case, we would have detected a pure molecular state. It is well known that the wave function and eigen energy in eqns.~\ref{wavefunction} and \ref{expectation value} account for the combined system of detector+molecule i.e. it is the eigen energy of a mixed state \cite{Bransden}. A detailed study of this aspect requires careful consideration of the inherent invasive nature of quantum detection \cite{Bransden}. This topic is of considerable interest in ongoing research \cite{Bayar,Katsnelsona} but lies beyond the scope of this paper.\\

\section{Conclusions}
\label{sec:Conclusions}
We have presented an {\it in situ} method to detect the electronic eigen energies in the afterglow of gas discharges at ambient conditions. The principle of operation relies on the observation and analysis of stochastic fluctuations in the static potential of a single electrode. By using this method we could derive the energy distribution of electrons and holes present near the surface of the electrode. In a helium discharge, the electron spectrum coincides with the electronic states in He~I. A free hole gas was observed in a environment of metastable nitrogen. The interaction with the electrode and excited state molecules surface was through the surface Auger process. The exchange of charge is directed through omnipresent free electron and free hole gases. Recombination of electrons and holes was found to be the fundamental underlying process of reduction and oxidation. At ambient conditions the lifetime of a hole in an electron density yields 4.4$\pm$0.2 fs whereas the lifetime of an electron in an hole density yields 10.8$\pm$0.5 fs. These measured lifetimes are in fair agreement with theoretical estimates of the processes which yield respectively 4.8 and 7.5 fs. Based on the observed asymmetry in the energy distributions of electron and hole ensembles we suggest that under conditions of thermal equilibrium the behavior of a hole is not equivalent to that of an electron vacancy.

\section*{Acknowledgements}
The authors thank H.C.W. Beijerinck for the discussions and his comments and H.L.M. Lelieveld for proof-reading the manuscript. This work was financially supported by the dutch Ministerie van Economische zaken: regeling EOS (Energie Onderzoek Subsidie) through contract LT-01044 and by the Commission of the European Communities, Framework 6, Priority 5 'Food Quality and Safety' through contract FP6-CT-2006-015710. 


\section*{Appendix}
\subsection{Metastable molecular nitrogen}
\label{nitrogen plasma}
\begin{figure*}[t]
\centering
\begin{tabular}{cc}
\includegraphics[width=60mm, angle=-90]{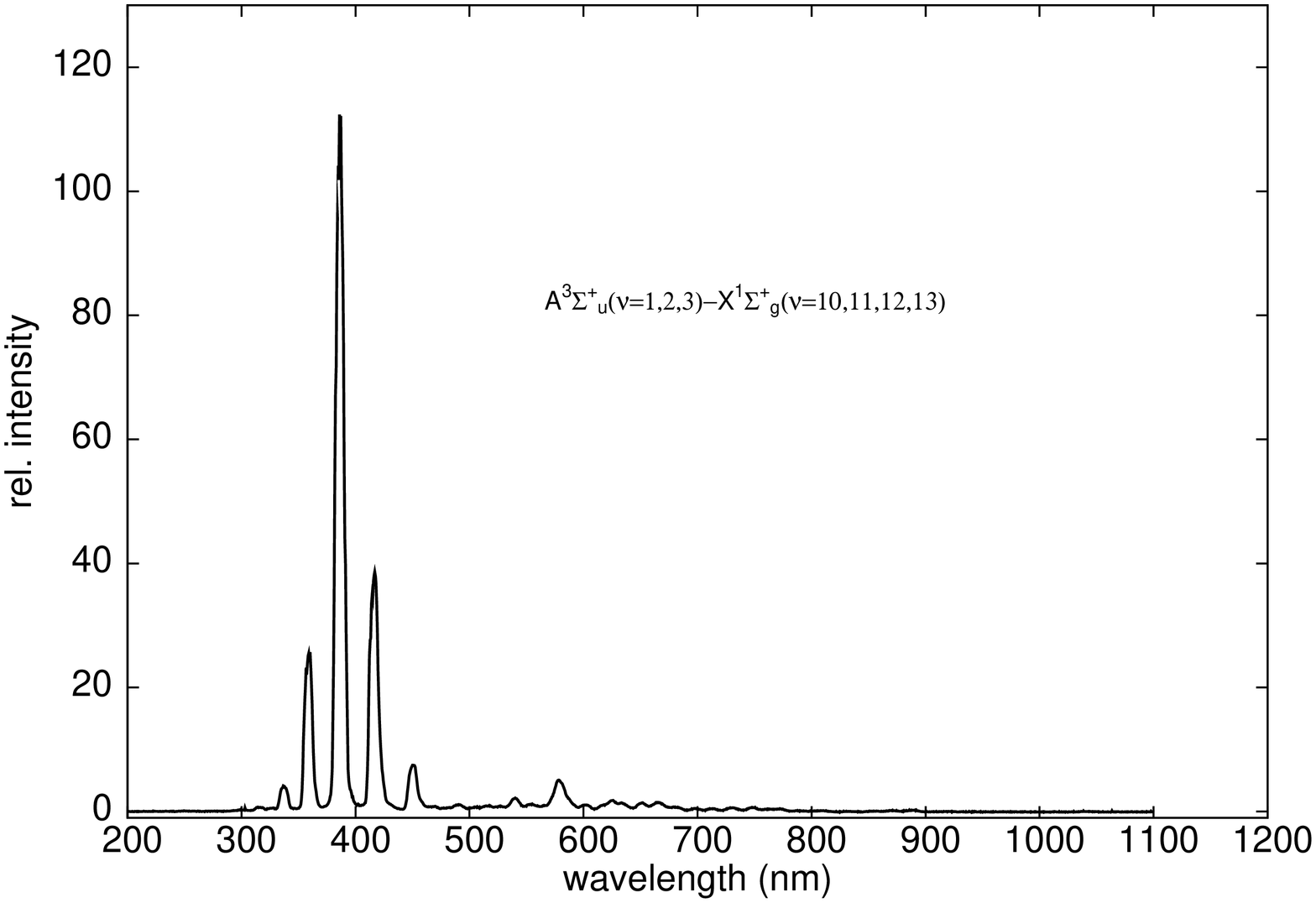}&\includegraphics[width=65mm, angle=-90]{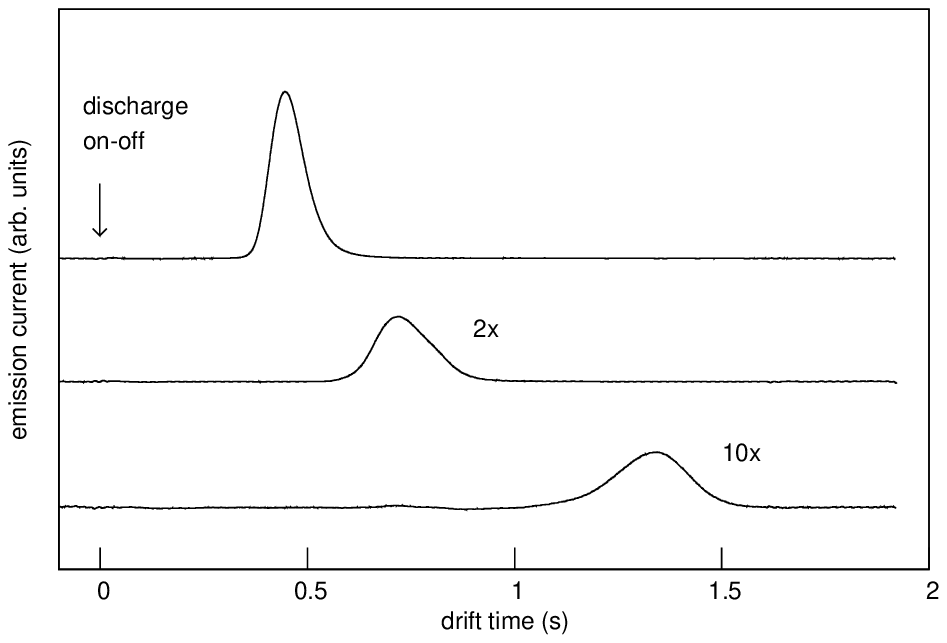}\\
(A)&(B)\\
\end{tabular}
\caption{\label{RTD} The molecular content of the nitrogen plasma jet was analysed by emission spectroscopy and lifetime measurements. (A) The characteristic forbidden $A^3\Sigma^+_u(\nu=1,2,3)$-$X^1\Sigma^+_g(\nu=10,11,12,13)$ Vegard-Kaplan emissions that were observed in the afterglow. (B) The lifetime of the excited state molecules, that was studied by Residence Time Distribution analysis using a 3 meter long drift tube, yields a lifetime on the order of 1 second.}
\end{figure*}

A plasma generated in a nitrogen environment at ambient conditions using a high voltage discharge contains a considerable amount of metastable nitrogen. Molecular metastable nitrogen $N_2$($A^3\Sigma^+_u)$, with an internal energy of 6.2eV and an estimated radiative lifetime of 1.9 sec \cite{Lofthus}, was previously observed at ambient conditions by \cite{Noxon}. We have verified that $N_2$($A^3\Sigma^+_u)$ was produced by our specific source by analysing the afterglow of the plasma using emission spectroscopy and lifetime measurements. Emission spectra were recorded perpendicular to the cold plasma jet using a fiber probe located 20mm above the nozzle. Light was analysed using an Avaspec 2048-2 UV-VIS spectrometer system supplied with a FC-UV400-2 fiber, UA-178-1100 grating and a Sony 2048 DUV detector board. We observed light emissions from the $A^3\Sigma^+_u(\nu=1,2,3)$-$X^1\Sigma^+_g(\nu=10,11,12,13)$ system (Vegard-Kaplan) \cite{Noxon,Lofthus}. In addition we verified the lifetime of the excited state molecules present in the jet by Residence Time Distribution analysis. After connecting a drift tube (Tricolair hose 10x16mm, 3m in length) to the nozzle we found that the light emission of the afterglow was extended over the entire length of the tube. The lifetime of the metastable fraction was analysed based on the survival of fraction exited state species detected at the end of the drift tube as a function of the Residence Time (RTD analysis). The excited state molecules were produced in bunches by periodically switching the high voltage of the discharge off-on-off with a 10\% duty cycle. We detected the relative numbers and the distribution of arrival times of the surviving excited state fraction using a Faraday cup (aluminia grid 20x20mm, 3mm mesh, 70\% open) located at the outlet of the tube. The emission current was detected using a Keithley 610C electrometer as a function of the time delay between the pulsing of the discharge and the arrival of excited state molecules at the detector. The gas flow was adjusted in the range of 0.1-0.3 L/s at laminair air flow conditions to control the residence time of excited state nitrogen in the drift tube. Typical residence time distributions and a corresponding emission spectrum are plotted in figure~\ref{RTD}. The spectroscopic data in conjunction with the observed lifetime of the excited state nitrogen is attributed to the presence of $N_2$($A^3\Sigma^+_u)$.

\begin{figure}
\center
\includegraphics[angle=0,height=50mm]{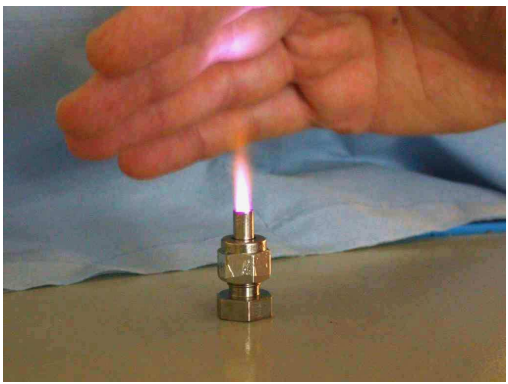}
\caption{\label{picture jet} The jet of excited state nitrogen (cold plasma) is based on a high voltage discharge operated at ambient conditions according to \cite{Strutt,Noxon,Anders}. The throughput of the carrier gas is typically 0.2~L/s. The purity of the nitrogen was 99.5$\pm$0.2\% which was determined using a Varian Micro MC-CP-2002 gaschromatography. The temperature of the cold plasma flame, measured using a mercury thermometer, was 40$\pm$2$^\circ$C.}
\end{figure}

Cold plasma's based on electric discharges in nitrogen (figures~\ref{set-up} and \ref{picture jet}) that operate under ambient conditions have been studied since 1916 \cite{Strutt,Noxon,Anders,Kieft}. Spectroscopic studies in the afterglow \cite{Lofthus} of nitrogen have revealed the presence of metastable molecular nitrogen. We could confirm that the giant afterglow \cite{Noxon} in nitrogen is due to metastable $N_2$ by combining emission spectroscopy with lifetime measurements (figure~\ref{RTD}). It appears that $N_2$(A), which has an internal energy of $E_A$=6.2eV, can exist at concentration of approximately 1 ppm in a thermal equilibrium with ground state $N_2$ at ambient temperatures (figures~\ref{picture jet} and \ref{RTD}). This is remarkable, as under these experimental conditions $N_2$(A) is able to survive more than $10^9$ exchange reactions $N_2(A)+N_2 \rightarrow N_2+N_2$(A) before detection. One could argue, in order for the second law of thermodynamics to hold, that only a fraction of magnitude $n_A/n_o=e^{-E_o/E_A}=e^{-kT/E_A}$=2.0$\cdot 10^{-100}$ should be found in the present experiments. However, this argument is incorrect. The reason why an extreme afterglow can be observed, even at ambient conditions, is two fold. First, the lowest exited state $A^3\Sigma^+_u$ is a triplet state whereas the ground state is a singlet. Radiative transitions between triplet and singlet are forbidden by spin selection rules and therefore unlikely to occur. Secondly, $N_2$(A) is stable in a $N_2$ environment as a result of the symmetry of the exchange reaction $N_2(A)+N_2 \rightarrow N_2+N_2$(A) which additionally excludes de-excitation by means of two body interactions. Similar to the nuclear Overhauser effect \cite{Reif} observed in spin polarized gases \cite{Albert,Ebert} the relaxation of $N_2$(A) is only by virtue of higher order de-excitation processes: the system is inherently non-stationary. Thus, despite that the fraction of $N_2(A)$ has completely thermalised, no true stationary equilibrium is reached in the afterglow. Thermodynamical laws simply do not hold \cite{Evans} under these 'extreme' conditions, thereby resolving the paradox. We conclude that $N_2$(A) is stable in a $N_2$ environment and that its decay is primarily through radiative emissions of forbidden light (Vegard-Kaplan) \cite{Lofthus}. 

\subsection{Voltage probe}
\label{sec:Probe}
In order to reduce the load capacitance and simultaneously increase the input impedance of our voltage probe (figure~\ref{set-up}) we used a SMD resistor (MMB0207 Vishay BC components) of 8.2MOhm//1pF (the residual stray capacitance across the SMD resistor yields $\sim$1pF) in series with the 1 MOhm//12pF input impedance of the recording oscilloscope (Tektronix TDS-3052B). Note that this probe operates according to the scheme of a nearly matched, high frequency voltage probe \cite{Regtien}. However, a connection cable between the high impedance resistor and the 1MOhm input resistance was avoided in our set-up. In this way the load capacity could be reduced to a pF level. This high impedance, low capacity probe allows the detection of voltage fluctuations at relatively small charging currents. The voltage readings of our probe were calibrated against a 30.0$\pm$0.1 VDC reference. Alternatively, we used a Tektronix P6139A (10:1) voltage probe. The P6139A has an input impedance of 10MOhm//8.0pF. Its relative high load capacitance is due to the compensation of the 0.7m long cable between the tip and the connector. The free hole and free electron gases reported in this work were observed using either the 8.2MOhm//1pF probe or the P6139A probe. The voltage reference of the recording oscilloscope and probe was biased using an external DC power supply (Hewlett Packard 412B) as indicated in figure~\ref{set-up}. 

\subsection{Data reduction and fitting routine} 
The fluctuations at a fixed bias level were analysed by building a histogram (bin sizes 200 meV or 25 meV) of the observed voltages. It should be obvious that the choice of the bin size when building a histogram affects the resolution of the observed energy distribution. A time series $V(t)$ typically contains 10.000 readings and series were recorded at a rate of 1kHz. Each data point $V(t)$ was converted to a sample of the energy $E_i$ using $E_i=-qV(t_i)$. The energy distributions thus obtained were fitted by means of a non-linear least-squares routine using the function:

\begin{eqnarray}
\frac{dn(E)}{dE}=
\begin{cases}
B_e \sqrt{E} \left[exp\left[\beta_e(E-E_s)\right]+1\right]^{-1} & E>0 \\
B_h \sqrt{-E} exp\left[\beta_h E\right] & E<0 
\end{cases}
\label{lineshape}
\end{eqnarray}

where $\beta_e=kT_e$, $E_{s}$ and an overall scaling factor $B_e$ are the fitting parameters for the electron gas and $\beta_h=kT_h$ and an overall scaling factor $B_h$ are the fitting parameters for the hole gas. To account for a mutual offset in the energy scale, the fit function in eqn.~\ref{lineshape} was allowed to shift (using an additional fitting parameter). Typically, the offset compensation yields approximately 120 meV. By using this fitting routine estimates were obtained for the temperature of the electron gas, the temperature of the hole gas and for the Fermi energy of the electron gas. Based on the information contained in all our data it is concluded that the hole gas corresponds to a fermion gas in the limiting case of high temperatures and low density. The Fermi-Dirac distribution factor in the fitting routine was therefore reduced to a Maxwell-Boltzmann distribution. This behavior is different for the electron gas. The electron energy distribution could be fitted only when fully accounting for the Fermi-Dirac statistics: the electron energy distribution has a non-zero shift in the Fermi energy. At negative bias $E_s$ typically yields 400 meV at a temperature of $kT_e$=150 meV, implying that we deal with the non-degenerate case (high density, low temperature limit) of a free electron gas. The estimates of the fitting parameters obtained by this fitting routine reproduces well if data are acquired for a fixed position of the electrode (25$\pm$5 mm) in the plasma flame.


\begin{thebibliography}{99}
\bibitem{Cagnon} Gagnon E, Ranitovic P, Tong X M, Cocke C L, Murnane M M, Kapteyn H C and SandhuSoft A S, {\it Science} {\bf 316}, 1374-1378 (2007). 
\bibitem{Niehaus}Hotop H and Niehaus A, {\it Z. Phys. } {\bf 228(1)}, 68-88 (1969).
\bibitem{Ohno}Ohno K, Mutoh H and Harada Y, {\it J. Am. Chem. Soc.} {\bf 105(14)}, 4555-4561 (1983).
\bibitem{Lemay}Zevenbergen M A G, Krapf D, Zuiddam M R, Lemay S G, {\it Nano Letters} {\bf 7}, 384-388 (2007).
\bibitem{Heller}Heller I, Kong J, Williams K A, Dekker C and Lemay S G, {\it J. Am. Chem. Soc.} {\bf 128(22)}, 7353-7359 (2006).
\bibitem{Frenken}Hendriksen B L M and Frenken J W M, {\it Phys. Rev. Lett.} {\bf(89)4}, 046101 (2002).
\bibitem{Reif} Reif F, {\it Fundamentals of Statistical and Thermal Physics} (McGraw-Hill, 1965).
\bibitem{Ashcroft}Ashcroft N W and Mermin N D, {\it Solid State Physics} (Saunders College, Philadelphia, 1976).
\bibitem{Hotop}Hotop H, in {\it Experimental methods in the physical sciences} {\bf 29B}, Dunning F D Ed.(Academic Press, San Diego, 1996).
\bibitem{Bransden}Bransden B H, Joachain C J, {\it Introduction to Quantum Mechanics} (Longman, Essex, (1989).
\bibitem{Jackson}Jackson JD, {\it Classical Electrodynamics} (Wiley, 1975) [2nd edition]
\bibitem{Strutt}Strutt R J, {\it Proc. Roy. Soc.} {\bf A92}, 438 (1916).
\bibitem{Noxon}Noxon JF, {\it J Chem. Phys.} {\bf 36(4)}, 926-940 (1962). 
\bibitem{Anders}Anders A and Kuhn M {\it Rev. Sci. Instr.} {\bf(69)3}, 1340-1344 (1998).
\bibitem{Kieft}{Kieft I E, vd Laan EP and Stoffels E }{\it New Journal of Physics}{\bf 6(149)}(2004).
\bibitem{Lofthus}Lofthus A, Krupenie P H, {\it J. Phys. Chem. Ref. Data} {\bf 6}, 113 (1977).
\bibitem{Regtien}Regtien P P L, {\it Instrumentele Elektronica} (Delftsche Uitgeversmaatschappij, 1986).
\bibitem{Nist}Ralchenko Y, Jou F C, Kelleher D E, Kramida A E, Musgrove A, Reader J, Wiese W L and Olsen K, {\it NIST Atomic Spectra Database} (2007). [version 3.1.2]
\bibitem{Bayar}Bayer M, Hawrylak P, Hinzer K, Fafard S, Korkusinski M, Wasilewski Z W, Stern O and Forchel A, {\it Science} {\bf 291}, 451-453 (2001).
\bibitem{Albert}Albert M S, Cates G D, Driehuys B, Happer W, Saam B, Springer Jr CS and Wishnia A, {\it Nature} {\bf 370}, 199-201 (1994).
\bibitem{Ebert} Ebert M, Grossmann T, Heil W, Otten W E, Surkau R, Leduc M, Bachert P, Knopp M V, Schad LR and Thelen M, {\it Lancet} {\bf 347} 1297-1299 (1996).
\bibitem{Heer} de Heer W A, Chatelain A and Ugarte D,{\it Science} {\bf 270}, 1179-1180 (1995).
\bibitem{Yang} Yang H Y, Lau S P, Yu S F, Huang L, Tanemura M, Tanaka J, Okita and Hng H H,{\it Nanotechnology} {\bf16}, 1300-1303 (2005).
\bibitem{Bonard} Bonard J M, Kind H, Stockli T, Nilsson L,{\it Solid State Electronics} {\bf45}, 893-914 (2001).
\bibitem{Evans} Evans D J and Searles D J,{\it Phys. Rev.} {\bf E50}, 1645 (1994). 
\bibitem{Slegel} Schlegel H G, {\it  General Microbiology} (Cambridge University Press, 1993)
\bibitem{Feynman} Feynman R P,{\it  Quantum Electrodynamics} (Westview Press, 1997)
\bibitem{Weinberg} Weinberg S,{\it The Quantum Theory of Fields} (Cambridge University Press, 2001)
\bibitem{Katsnelsona} Katsnelson M I and Novoselov KS,{\it Solid State Commun.} {\bf 143(1-2)}, 3-13 (2007).

\end{thebibliography}
\end{document}